\documentclass[a4paper,11pt]{article}
\usepackage{pos}
\renewcommand{\logo}{\relax}

\usepackage{siunitx}
\usepackage[capitalise]{cleveref}
\usepackage{xfrac}
\usepackage{booktabs}

\title{Cabibbo angle anomalies and a global fit to vector-like quarks}

\author*[a]{Matthew Kirk}

\affiliation[a]{Departament de Física Quàntica i Astrofísica (FQA) \& Institut de Ciències del Cosmos (ICCUB), Universitat de Barcelona (UB), Barcelona, Spain}

\emailAdd{mjkirk@icc.ub.edu}

\abstract{
The most recent determinations of $V_{ud}$ from superallowed beta decays lead to a discrepancy when compared to the value implied by mesonic CKM measurements combined with CKM unitarity.
On top of this, improved precision in lattice QCD calculations have revealed another discrepancy between the $V_{us}$ determinations from kaon and pion semi-leptonic decays. 
The combination of these can be referred to as the Cabibbo angle anomaly, which we find has a significance of around \qty{3}{\sigma}.
After summarising the current state of these issues, I will talk about new physics models that modify semi-leptonic decays as potential explanations, and why vector-like quarks in particular appear the most promising candidates.
I will then discuss the results of a global fit to various vector-like quark models, and how other constraints are important in determining the most likely explanation.
Finally I will touch on future experiments that could shed further light on the situation.
}

\FullConference{21st Conference on Flavor Physics and CP Violation (FPCP 2023)\\
 29 May - 2 June 2023\\
 Lyon, France\\}

\begin{document}

\renewcommand{\hookAfterAbstract}{%
\par\bigskip
\textsc{ArXiv ePrint}:
 \href{https://arxiv.org/abs/2212.06862}{2212.06862}
}
\maketitle

\section{CKM Matrix and Unitarity}

\begin{figure}
\centering
\includegraphics[width=0.7\textwidth]{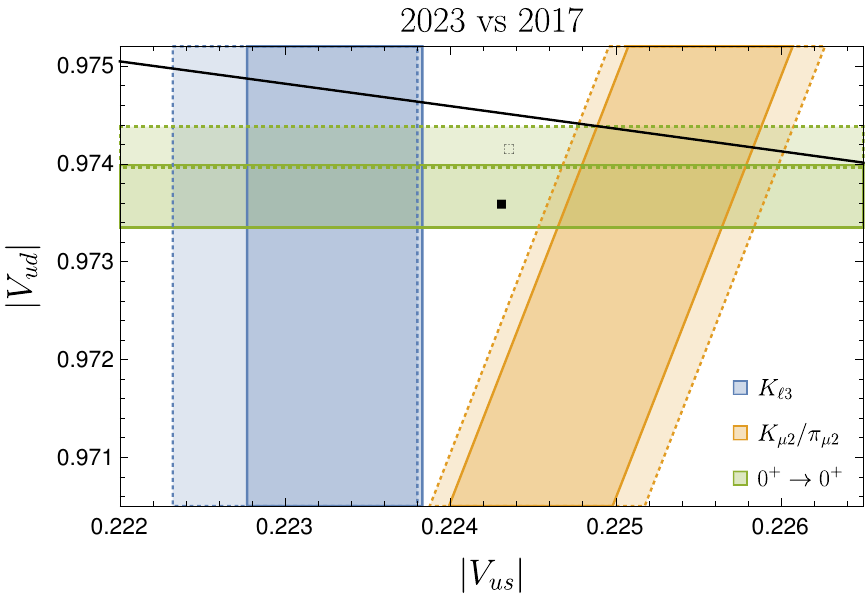}
\caption{Status of the three best determinations of $V_{ud}$ and $V_{us}$ in 2023 (solid) and 2017 (dashed).
Bands represent \qty{1}{\sigma} regions, and the square indicates the best fit point.}
\label{fig:Vud_Vus_SM_2023_vs_2017}
\end{figure}

In the Standard Model (SM), the CKM matrix is unitary by construction (since it is the product of two unitary matrices from the diagonalisation of the SM quark Yukawa matrices).
Combined with the freedom to rephase the quark fields, the nine complex elements of the CKM matrix are related to just four independent parameters, and there are therefore many relationships between CKM elements which we can use to probe this ``unitarity prediction''.
Amongst others, we focus on the ``first row unitarity'' relation
\begin{equation}
|V_{ud}|^2 + |V_{us}|^2 + |V_{ub}|^2 \equiv 1 \,.
\end{equation}
Because $|V_{ub}|^2$ is so small ($\sim \num{1e-5}$ \cite{Charles:2004jd,CKMfitter:Spring2021}) we can neglect its contribution, which is smaller than the current errors on the other two terms, and consider the simpler two family relationship
\begin{equation}
\label{eq:first_row_unitarity}
|V_{ud}|^2 + |V_{us}|^2 \overset{?}{=} 1 \,,
\end{equation}
where the question mark indicates that we are going to test whether this relation holds.\footnote{The SM does not predict the individual values, which is part of the wider flavour problem of the SM.}
At the moment, there are three determinations of $V_{ud}$ and $V_{us}$ that are significantly more precise than the others:
\begin{enumerate}
    \item $V_{ud}$ from superallowed beta decays (green, $0^+ \to 0^+$),
    \item $V_{us}$ from three-body semi-leptonic kaon decays (blue, $K_{\ell 3}$),
    \item $V_{us} / V_{ud}$ from the ratio of kaon to pion two-body leptonic decays (orange, $K_{\mu 2} / \pi_{\mu 2}$),
\end{enumerate}
which we show in \cref{fig:Vud_Vus_SM_2023_vs_2017} (other less precise determinations can be found in \cite{Crivellin:2022rhw}).
From the figure, we can see two things: 1) that the data do not agree well amongst themselves, and 2) that the best fit point does not lie on the SM prediction line (in black).%
\footnote{We comment that in a two family model the CKM reduces to the $2 \times 2$ Cabibbo matrix with a single free parameter, the Cabibbo angle, from where this discrepancy gets the name ``Cabibbo Angle Anomaly'', since each determination gives a different value for this parameter.}
In fact we find that the current tension with the unitarity prediction of \cref{eq:first_row_unitarity} is around \qty{3}{\sigma} \cite{Crivellin:2022rhw}, and this tension is what we refer to as the ``Cabibbo Angle Anomaly'' (CAA).

\section{Recent changes to \texorpdfstring{$V_{ud}$}{Vud} and \texorpdfstring{$V_{us}$}{Vus}}

This anomaly has developed over the last 5 or so years, as we see from the dotted lines in \cref{fig:Vud_Vus_SM_2023_vs_2017} that represent the situation in 2017.
We see that there was a smaller tension between the kaon and pion decay results, as well as a somewhat higher value for $V_{ud}$ from beta decays, all of which combined gave a best fit point that was in much more agreement with the SM unitarity prediction.
So what happened?
There have been two main developments, which we will briefly discuss.

\subsection{Lattice QCD}
Several new lattice QCD calculations in recent years have improved our knowledge of the relevant decay constants and form factors.
Comparing the FLAG $N_f = 2+1+1$ averages from 2017 \cite{Aoki:2016frl} (using \cite{Carrasco:2014poa,Dowdall:2013rya,FermilabLattice:2014tsy,Carrasco:2016kpy,Bazavov:2013maa}) and 2023 \cite{FlavourLatticeAveragingGroupFLAG:2021npn} (using \cite{Carrasco:2016kpy,FermilabLattice:2018zqv,Bazavov:2017lyh,Dowdall:2013rya,Carrasco:2014poa,Miller:2020xhy,ExtendedTwistedMass:2021qui}):
\begin{align}
\frac{f_K}{f_\pi} &= \num{1.1930 \pm 0.0030} \to \num{1.1934 \pm 0.0019} \,,
\\
f_+^K (0) &= \num{0.9706 \pm 0.0027} \to \num{0.9698 \pm 0.0017} \,,
\end{align}
we see that errors on both have reduced by around a factor of $\sfrac{1}{3}$, while the central values are almost identical.
This improvement accounts for the shrunken error bars of the orange and blue bands between the 2017 and 2023 data in \cref{fig:Vud_Vus_SM_2023_vs_2017}, which made clear that there is some tension in those results alone.

\subsection{Nuclear and electroweak corrections}

The theoretical prediction for a particular nuclear beta decay requires various corrections as compared to the free quark level process $d \to u e \nu$.
Firstly there are the 1-loop electroweak (EW) corrections (which are often called universal corrections), primarily from a $\gamma W$ box diagram, and these have been the focus (and indeed were a spur) of discussion around the CAA.
The current average value \cite{Cirigliano:2022yyo} of several new calculations \cite{Seng:2018yzq,Seng:2018qru,Czarnecki:2019mwq,Seng:2020wjq,Hayen:2020cxh,Shiells:2020fqp}, is around three standard deviations larger than the value from 2005 \cite{Marciano:2005ec} which was used for many years.
This accounts for the vertical shift in the $0^+ \to 0^+$ band between the 2017 and 2023 data in \cref{fig:Vud_Vus_SM_2023_vs_2017}.
Following on from these calculations, the community also reassessed various other corrections associated with nuclear structure \cite{Gorchtein:2018fxl} and isospin breaking terms \cite{Miller:2008my,Miller:2009cg,Condren:2022dji,Seng:2022epj,Crawford:2022yhi}, and typically a larger uncertainty is now assumed (which in fact is now dominant), that can be seen in the increased width of the $0^+ \to 0^+$ band.

\section{Vector-like quarks as the best BSM explanation}

Since these CKM elements are extracted from semi-leptonic decays, there is a wide range of BSM models that can alter these processes and therefore potentially affect the extraction.
In the literature, studies have been made of leptoquarks \cite{Crivellin:2021egp,Crivellin:2021bkd}, $W'$s \cite{Kirk:2020wdk,Capdevila:2020rrl}, vector-like leptons \cite{Crivellin:2020ebi,Kirk:2020wdk,Coutinho:2019aiy,Endo:2020tkb}, vector-like quarks (VLQs) \cite{Belfatto:2019swo,Belfatto:2021jhf,Branco:2021vhs,Crivellin:2022rhw,Belfatto:2023tbv}, charged scalars \cite{Crivellin:2020klg,Crivellin:2020oup,Marzocca:2021azj,Felkl:2021qdn}, or $Z'$s \cite{Buras:2021btx}.
Of these, VLQs seem the most promising, since they offer the potential to generate right-handed charged currents in the quark sector, which alter three-body and two-body decays differently, and therefore can reduce the tension between the $K_{\ell 3}$ and $K_{\mu 2} / \pi_{\mu 2}$ CKM results.
(In fact, VLQs are one of only two single particle extensions of the SM that can do this -- the only other one is a charged $SU(2)_L$ singlet vector boson with flavour changing couplings \cite{deBlas:2017xtg}.)

Vector-like quarks can be defined as new coloured fermions whose left and right handed components have the same representation under the SM gauge group.
There are seven possible fields of this type which couple to the SM quarks at tree level, which are shown in \cref{tab:vlq_reps}.
\begin{table}
\centering
\begin{tabular}{@{}l |  c c c c c c c @{}}
\toprule
& $U$ & $D$ & $Q$ & $Q_5$ & $Q_7$ & $T_1$ & $T_2$ \\
\midrule
$SU(3)_C$ & 3 & 3 & 3 & 3 & 3 & 3 & 3 \\
$SU(2)_L$ & 1 & 1 & 2 & 2 & 2 & 3 & 3 \\
$U(1)_Y$ & \sfrac{2}{3} & \sfrac{-1}{3} & \sfrac{1}{6} & \sfrac{-5}{6} & \sfrac{7}{6} & \sfrac{-1}{3} & \sfrac{2}{3} \\
\bottomrule
\end{tabular}
\caption{Possible representations of VLQs under the SM gauge group.}
\label{tab:vlq_reps}
\end{table}
Of these, the $SU(2)$ singlets and triplets will alter the left-handed $W$ (and $Z$) couplings after being integrated out, while the $SU(2)$ doublets alter right-handed couplings.
In fact, only the $Q$ VLQ generates a right-handed $W$-$u$-$d$ type interaction, and it is this field we focus on.

To quickly dispense with the other options, we note that the $T_1$ and $T_2$ triplets generate the wrong sign modification of $W$ couplings to reduce the CAA tension.
On the other hand, the $U$ and $D$ singlets give the opposite sign modification, but $SU(2)$ invariance can lead to new flavour-changing $Z$ currents (depending on the assumed couplings) which are well constrained by either $D$ or $K$ meson mixing, while even the modified flavour conserving $Z$ interactions are strongly bounded by electroweak precision observables (EWPO) and low-energy parity violation measurements.

\begin{figure}
\centering
\includegraphics[width=\textwidth]{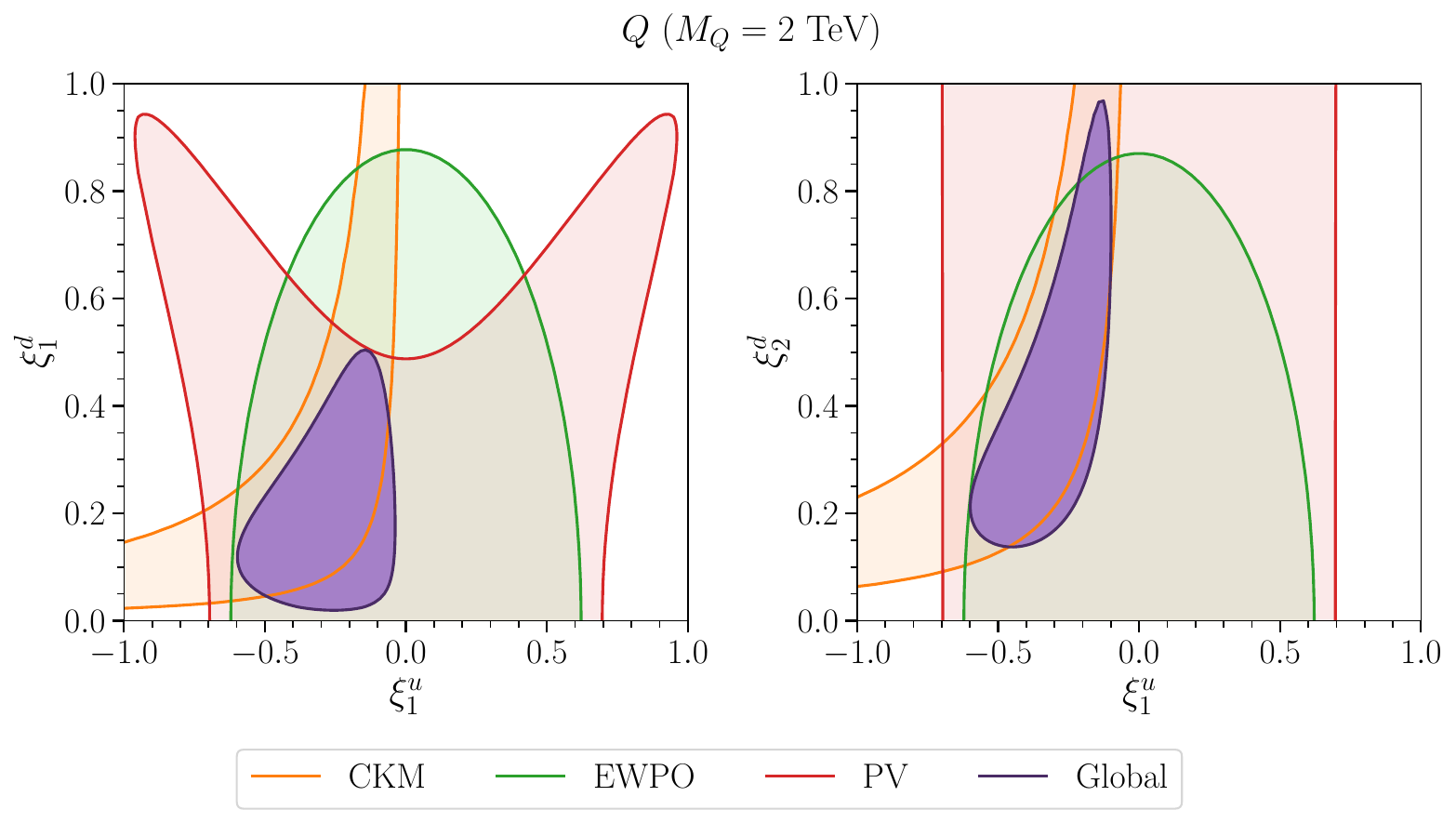}
\caption{Our global fit to the $SU(2)$ doublet vector-like quark $Q$, assuming it has couplings only to right-handed up and down quarks on the left, or only right-handed up and strange quarks on the right.
See our original paper \cite{Crivellin:2022rhw} for precise details of the interaction Lagrangian and the observables contained within our global fit.}
\label{fig:Q1_VLQ_fit}
\end{figure}

Returning to our star pupil, the $SU(2)$ doublet $Q$, we find that meson mixing bounds are either totally absent or highly suppressed, even at 1-loop, while the EWPO limits for right-handed modified $Z$ couplings are somewhat weaker than in the LH sector.
Here parity violation (PV) observables are equally strong, as we see from our global fit in \cref{fig:Q1_VLQ_fit}.
Crucially, we find that the favoured region to explain the CAA (shaded orange and denoted ``CKM'') is almost totally unconstrained by the other bounds, and that the global best fit region (purple) nicely overlaps, allowing for a full explanation.

\section{Future experiments}
In light of the ongoing developments in beta decays, another independent and theoretically clean measurement of $V_{ud}$ would be extremely welcome to shed light on the situation.
Pion beta decay ($\pi^+ \to \pi^0 e \nu$) is exactly that -- a theoretically very well understood decay which can provide a new and competitive determination of both $V_{ud}$, as well as $V_{us} / V_{ud}$ in conjunction with the three-body kaon decay \cite{Czarnecki:2019iwz}.
The PIONEER experiment \cite{PIONEER:2022alm,IwamotoFPCP2023} aims to measure this branching ratio with a precision of less than \qty{0.2}{\percent} in phase II (a factor of three improvement compared to the current status), which would be sufficient for this goal.%
\footnote{We also note that PIONEER will first measure the lepton flavour universality ratio $\pi \to e \nu / \pi \to \mu \nu$, which would help test other potential explanations of the CAA such as vector-like leptons.}

Furthermore, the tensions in the mesonic extractions could be sharpened (or lessened) with a new measurement of the $K \to \mu \nu$ decay mode.
This mode was most recently measured by KLOE in 2008, and currently there exists some tension amongst that and the previous data \cite{ParticleDataGroup:2022pth}.
In the analysis of \cite{Cirigliano:2022yyo}, the authors demonstrated that the NA62 experiment could measure the ratio $K \to \pi \ell \nu / K \to \mu \nu$, and a short two week run would likely provide sufficient data to increase the CAA tension to almost \qty{4}{\sigma} if the measured value is two standard deviations below the current average (or reduce the tension to close to \qty{1}{\sigma} if the shift was in the other direction).

\section{Summary}
We have discussed the current state of the emerging Cabibbo Angle Anomaly, where improvements to lattice QCD and beta decay calculations have lead to a roughly \qty{3}{\sigma} discrepancy with the SM.
Amongst a wide array of possible BSM explanations, we have shown that, in particular, an $SU(2)$ doublet vector-like quark shows the most promise.
Finally we discussed how new data could sharpen this tension, and further illuminate this interesting discrepancy.

\section*{Acknowledgements}
I would like to thank the organisers of FPCP2023 for the chance to present, and a very enjoyable conference overall.
I would also like to acknowledge Andreas Crivellin, Teppei Kitahara, and Federico Mescia as my co-authors of \cite{Crivellin:2022rhw}, for the pleasant collaboration that lead to that work and for their comments on this article.

\bibliographystyle{JHEP}
\bibliography{refs.bib}

\end{document}